\documentclass[12pt]{iopart}

\usepackage{iopams}
\usepackage{graphicx}

\newcommand{\Align}[1]{\begin{eqnarray}#1\end{eqnarray}}

\newcommand{\Cases}[1]{
\left \{
\begin{array}{ll}
#1
\end{array}
\right.
}

\newcommand{\Description}[1]{\begin{description}#1\end{description}}

\newcommand{\bfm}[1]{{\mathbf #1}}
\newcommand{\bbm}[1]{{\mathbb #1}}

\newcommand{\eqref}[1]{(\ref{#1})}

\newcommand{\beq}{\begin{equation}}
\newcommand{\eeq}{\end{equation}}
\newcommand{\ee}[1] {\label{#1} \end{equation}}
\newcommand{\bea}{\begin{eqnarray}}

\newcommand{\eea}{\end{eqnarray}}

\newcommand{\brac}[1]{ \left( #1 \right) }
\newcommand{\braC}[1]{ \left\{ #1 \right\} } 

\newcommand{\braccc}[1]{ \left[ #1 \right] }

\newcommand{\ie}{{\em i.e.}\ }

\newcommand{\Verb}[1]{\verb|#1|}

\begin{document}
\title{Meeting time distributions in Bernoulli systems}
\author{A. Akaishi$^1$, M. Hirata$^2$, K. Yamamoto$^3$ and A. Shudo$^1$}
\address{$^1$Department of Physics, Tokyo Metropolitan University, Tokyo 192-0397, Japan}
\address{$^2$Department of Mathematics, Tokyo Metropolitan University, Tokyo 192-0397, Japan}
\address{$^3$Department of Mathematics, Tokyo Institute of Technology, Tokyo 152-8551, Japan}

\ead{akaisi-akira@ed.tmu.ac.jp}

\begin{abstract}
Meeting time is defined as the time for which two orbits approach each other within distance $\epsilon$ in phase space.
We show that the distribution of the meeting time is exponential in $(p_1,\cdots,p_k)$-Bernoulli systems.
In the limit of $\epsilon\to0$, the distribution converges to $\exp(-\alpha\tau)$, 
where $\tau$ is the meeting time normalized by the average.
The exponent is shown to be $\alpha=\sum_{l=1}^{k}p_l(1-p_l)$ for the Bernoulli systems. 
\end{abstract}

\maketitle

%%%%%%%%%%%%%%%%%%%%%%%%%%%%%%%%
\section{Introduction}

In hyperbolic systems, two nearby trajectories separate exponentially in time, 
which is characterized by Lyapunov exponents, 
the rate of exponential growth of \emph{infinitesimal} initial deviations \cite{LL:book:83,Gas:book:98,Sko:LNP:10}.
On the other hand, the Poincar\'e recurrence theorem tells us 
that a trajectory starting at a compact \emph{finite} size region in phase space returns 
to the initial region infinitely-many times \cite{CFS:book:82,Ott:book:02}.
These basic properties characterize complex nature of chaotic systems.

The recurrence time statistics
is a reliable tool to measure correlations of chaotic trajectories \cite{CS:PRL:99,ZPRTK:PRE:07,ZTRK:Cha:07,JKA:EPL:10}:
if a system is purely chaotic, 
the recurrence time distribution is rigorously shown to be exponential in the limit of small recurrence regions \cite{Hir:ETDS:93,HSV:CMP:99}.
In mixed phase space 
where chaotic components and regular islands coexist, 
the recurrence time obeys a power-law distribution and its exponent is predicted to be universal 
\cite{CS:PRL:99,WHK:PRL:02,CK:PRL:08,Ven:PRL:09}, 
although controversial issues still remain on the existence of the universality and 
the ambiguity of how one determines proper initial recurrence regions.

In this paper, by analogy with the recurrence time, we will 
consider time to return to not fixed but moving regions.
Consider a region which is a neighborhood of a given orbit 
and an orbit starting at the region.
One expects that two nearby orbits which are initially located at a small but finite-length distance separate 
and, after a while, approach each other within the initial length, possibly infinitely-many times.
Statistical properties of time intervals for which two orbits 
come close to each other
must contain relevant information to chaotic 
dynamics. 
In the present study we refer to it as meeting time.
Although the meeting time is based on the simple idea, 
to the author's knowledge, it is not closely studied 
as far.
In particular, 
we investigate the meeting time in fully chaotic systems.

In \sref{sec:2} we define the meeting time distributions for maps with 
compact
phase space.
\Sref{sec:3} describes $\brac{1/k,\cdots,1/k}$-Bernoulli systems, 
for which 
one can obtain the 
meeting time distribution rigorously.
In \sref{sec:4} we discuss the meeting time in general Bernoulli systems,
and concluding remarks are given in \sref{sec:5}.

%%%%%%%%%%%%%%%%%%%%%%%%%%%%%%%%
\section{A definition of the meeting time}
\label{sec:2}

Let $X$ be a phase space and 
$F$ be a map on $X$ and, for $x,x'\in X$, $d(x,x')$ be a distance.
Then, for $x_1,x_2\in X$ and $\epsilon>0$, a sequence of 
times at which the two trajectories 
approach each other within a distance $\epsilon$ is given as
%%%%%%%%
\Align{
n^{(0)}_{\epsilon}&=\inf\braC{n\in\bbm{N}~|~d\brac{F^{n}(x_{1}),F^{n}(x_{2})}<\epsilon},\\
n^{(i)}_{\epsilon}&=\inf\braC{n\in\bbm{N}~|~d\brac{F^{n}(x_{1}),F^{n}(x_{2})}<\epsilon,n^{(i-1)}_{\epsilon}<n}.
}
%%%%%%%%
The meeting time is defined as the time interval between these ``meetings'': 
%%%%%%%%
\Align{
T^{(i)}_{\epsilon}=n^{(i+1)}_{\epsilon}-n^{(i)}_{\epsilon}.
}
%%%%%%%%
Then, the meeting time distribution is defined as
%%%%%%%%
\Align{
m_{\epsilon}(T)=\lim_{N\to\infty}\frac{1}{N}\#\braC{0\leq i <N~|~ T^{(i)}_{\epsilon}=T}.
\label{eq:def_dist}
}
%%%%%%%%
The meeting time depends, in general, on the choice of $x_1$ and $x_2$.
If $x_1$ and $x_2$ are both periodic orbits, the distribution is trivial.
(If simply $x_1$ and $x_2$ are both fixed points, $m_{\epsilon}(1)=1,\ m_{\epsilon}(T)=0\ (T\geq2)$ for $\epsilon\geq\epsilon_c$ 
where $\epsilon_c=d(x_1,x_2)$ and for $\epsilon<\epsilon_c$ the distribution cannot be defined.
More generally, if $x_1$ and $x_2$ have different periods, 
$L_1$ and $L_2$, 
$\sum_{T=1}^{T_{L_1,L_2}}m_{\epsilon}(T)=1,\ m_{\epsilon}(T)=0\ (T> T_{L_1,L_2})$ for certain $\epsilon$. 
Here $T_{L_1,L_2}$ is the least common multiple of $L_1$ and $L_2$.
)

In ergodic systems, for almost every $x_1$ and $x_2$ the distribution is independent of initial choice.
This is justified as follows:
as a function explicitly depending on $x_1$ and $x_2$, 
let us here rewrite the distribution as $m_{\epsilon}(T,x_1,x_2)$ and 
consider a sequence of the functions, namely $m_{\epsilon}(T,F^j(x_1),F^j(x_2))\ (j=0,1,\cdots)$.
By the definition of the distribution, 
it holds that $m_{\epsilon}(T,x_1,x_2)=\cdots =m_{\epsilon}(T,F^j(x_1),F^j(x_2))=\cdots$.
$^1$\footnote[0]{
$^1$This is justified as follows. Here we only show that $m_{\epsilon}(T,x_1,x_2)=m_{\epsilon}(T,F(x_1),F(x_2))$ and 
$m_{\epsilon}(T,F^j(x_1),F^j(x_2))=m_{\epsilon}(T,F^{j+1}(x_1),F^{j+1}(x_2))$ follows in the same way.
Instead of Eqs. (1)-(3), let us here express explicit dependence of $n^{(i)}$ and $T^{(i)}$ on $x_1$ and 
$x_2$ as $n^{(i)}_{\epsilon}(x_1,x_2)$ and $T^{(i)}_{\epsilon}(x_1,x_2)$. 
By the definition, $n^{(i)}_{\epsilon}(x_1,x_2)=n^{(i)}_{\epsilon}(F(x_1),F(x_2))-1$ holds 
and, therefore, the meeting time is given as $T^{(i)}_{\epsilon}(x_1,x_2)=T^{(i)}_{\epsilon}(F(x_1),F(x_2))$. 
This leads that $m_{\epsilon}(T,x_1,x_2)=m_{\epsilon}(T,F(x_1),F(x_2))$. 
}
For every $T$, $m_{\epsilon}(T,x_1,x_2)$ is an invariant function with respect to $x_1$ and $x_2$ under the map $F$, and, 
according to the ergodic theorem, is constant for almost every $x_1$ and $x_2$. 
(The exceptions are the cases for which $x_1$ and/or $x_2$ are periodic.)
We hereafter consider $m_{\epsilon}(T)$ for such $x_1$ and $x_2$.

We will discuss the meeting time by taking the limit $\epsilon\to0$.
Since smaller $\epsilon$ the meeting time becomes longer, 
the distribution should be taken into account with a proper normalization.
A plausible and canonical normalization 
would be an average,
%%%%%%%%
\Align{
\overline{T}_{\epsilon}=\lim_{N\to\infty}\frac{1}{N}\sum_{i=0}^{N-1}T^{(i)}_{\epsilon}.
\label{eq:average}
}
%%%%%%%%
The distribution normalized by the average meeting time is defined as
%%%%%%%%
\Align{
m'_{\epsilon}(\tau)=\lim_{N\to\infty}\frac{1}{N}\#\braC{0\leq i <N~|~ t^{(i)}_{\epsilon}=\tau},
}
%%%%%%%%
where $ t^{(i)}_{\epsilon}=T^{(i)}_{\epsilon}/\overline{T}_{\epsilon}$.

Remark that the meeting time 
could be regarded as recurrence time, more precisely
the recurrence for a cross product system of $(X,F)$: 
let $Y:=X\times X$ be a phase space and $G$ be a map on $Y$, that is $G(x_1,x_2):=(F(x_1),F(x_2))$.
The meeting time for a system $(X,F)$ is equivalent to 
the time to return to the diagonal region, $R_{\epsilon}=\{(x_1,x_2)~|~d(x_1,x_2)<\epsilon\}$, with respect to the system $(Y,G)$.
Thus, infinitely-many times meetings are 
proved to happen 
due to the Poincar\'e recurrence theorem.

%%%%%%%%%%%%%%%%%%%%%%%%%%%%%%%%
\section{Bernoulli systems}
\label{sec:3}

\subsection{$\brac{\frac{1}{2},\frac{1}{2}}$-Bernoulli systems}
\label{sec:3:1}

\begin{figure}
\begin{center}
\includegraphics{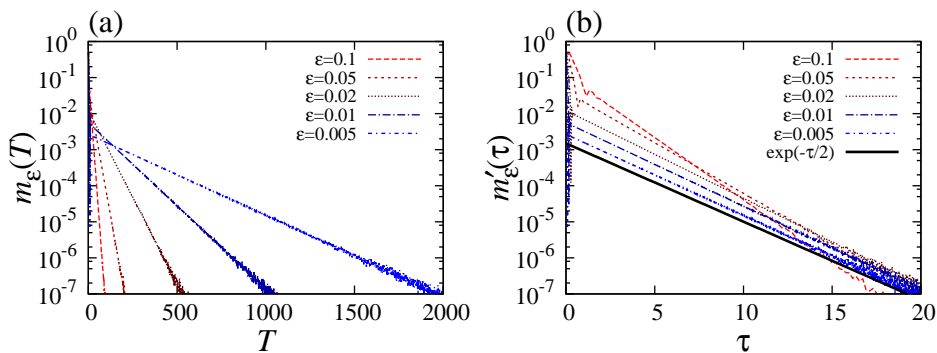}
\caption{
(a) The meeting time distributions for $\brac{\frac{1}{2},\frac{1}{2}}$-Bernoulli system.
(b) The same plot of (a) with the time normalized as $\tau=T/\overline{T}_{\epsilon}$ 
where the average $\overline{T}_{\epsilon}$ is numerically evaluated.
The black line represents the expected one by equation \eref{eq:dist_bernoulli2}.
}
\label{figure1a_1b}
\end{center}
\end{figure}

Let us consider binary symbolic dynamics:
let $X$ be a set of semi-infinite binary sequences, $\braC{0,1}^{\bbm{N}}$, 
and $F$ be a shift map on $X$.
We here assume $\brac{\frac{1}{2},\frac{1}{2}}$-Bernoulli measure.
For simplicity we put $\epsilon=\epsilon_r:=2^{-r}$ where $r$ is a positive integer.
In \ref{app:A}, we show that the meeting time distribution is given as,
%%%%%%%%
\Align{
m_{\epsilon_r}(T)&=
\Cases{
1/2&\brac{T=1}\\
0&\brac{2\leq T\leq r}\\
J^{(r)}_{T}2^{-T}&\brac{r+1\leq T}
}
,
}
%%%%%%%%
where $J^{(r)}_{T}$ is an integer 
generated by the following recurrence relation,
%%%%%%%%
\Align{
J^{(r)}_T&=\sum_{j=1}^{r}J^{(r)}_{T-j}\ \brac{r+1\leq T},
\label{eq:rec}
}
%%%%%%%%
and $J_{r+1}^{(r)}=1,\ J_{T}^{(r)}=2^{T-r-2}$ for $r+2 \leq T\leq 2r+1$.
One can easily check that $ \sum_{T=1}^{\infty}m_{\epsilon_r}(T)=1$, and
show that the average meeting time is given as
%%%%%%%%
\Align{
\overline{T}_{\epsilon_r}=\sum_{T=1}^{\infty}Tm_{\epsilon_r}(T)=2^{r}.
}
%%%%%%%%
The sequence given by the above recurrence relation \eqref{eq:rec} is called the 
{\it $r$th generalized Fibonacci number}.
The general solution of the $r$th generalized Fibonacci number is 
obtained by the roots of its characteristic polynomial
%%%%%%%%
\Align{
P_{r}(\lambda)&=\lambda^r-\lambda^{r-1}-\cdots-1.
}
%%%%%%%%
Let $\lambda_{(r)}$ denote the largest root of the 
characteristic
polynomial, \ie $P_{r}(\lambda_{(r)})=0$.
It is shown that $\lambda_{(r)}$ is real and its absolute value is greater than 1 
and the absolute value of the every other roots is smaller than 1 \cite{Mil:AMM:60}.
Since the polynomial has only one root whose absolute value is greater than 1, 
from general treatment of recurrence relations, 
one can see that the $r$th generalized Fibonacci number for large $T$ approximately 
behaves as 
%%%%%%%%
\Align{
J^{(r)}_T&\sim\lambda_{(r)}^T.
}
%%%%%%%%
Then, we obtain an asymptotic form of the distribution as
%%%%%%%%
\Align{
m_{\epsilon_r}(T)&\sim\brac{\frac{\lambda_{(r)}}{2}}^{T}.
}
%%%%%%%%
One can easily show $P_{r}(2)=1,\ P_{r}(\lambda)>1(\lambda>2)$ and thus $1<\lambda_{(r)}<2$ \cite{Mil:AMM:60}.
Since $P_{r}(2-2^{-r})={\cal O}(2^{-r})$, the root $\lambda_{(r)}$ for large $r$ is approximately given as
%%%%%%%%
\Align{
\lambda_{(r)}= 2-2^{-r}+{\cal O}(2^{-2r}).
}
%%%%%%%%
The normalized meeting time distribution is as follows:
%%%%%%%%
\Align{
m'_{\epsilon_r}(\tau)&\sim\brac{1-\frac{1}{2^{r+1}}}^{2^r \tau}.
}
%%%%%%%%
In the limit of large $r$, namely small $\epsilon_r$, the asymptotic form of the limiting distribution becomes 
%%%%%%%%
\Align{
\lim_{r\to\infty} m'_{\epsilon_r}(\tau)\sim \exp\brac{-\frac{\tau}{2}}.
\label{eq:dist_bernoulli2}
}
%%%%%%%%
\Fref{figure1a_1b} shows the meeting time distribution obtained 
by numerical simulations of $\brac{\frac{1}{2},\frac{1}{2}}$-Bernoulli systems.
As $\epsilon$ decreases, the normalized distribution converges to the exponential one of equation \eref{eq:dist_bernoulli2}.

%%%% figure %%%%
%%%% figure %%%%

\subsection{$\brac{1/k,\cdots,1/k}$-Bernoulli systems}
\label{sec:3:2}

The derivation 
made 
in the previous section can be easily extended to the case of $k$ symbols.
Let $X$ be $\braC{1,\cdots,k}^{\bbm{N}}$ and assume $\brac{1/k,\cdots,1/k}$-Bernoulli measure and put $\epsilon_r=k^{-r}$.
A combinatorial study allows us to derive the meeting time rigorously as well as the binary symbol case (see the last paragraph in \ref{app:A}).
The meeting time distribution for $\brac{1/k,\cdots,1/k}$-Bernoulli systems is given as
%%%%%%%%
\Align{
m_{\epsilon_r}(T)&=
\Cases{
1/k&\brac{T=1}\\
0&\brac{2\leq T\leq r}\\
J^{(r)(k)}_{T}k^{-T}&\brac{r+1\leq T}
}
,
}
%%%%%%%%
where $J^{(r)(k)}_{T}$ is an integer that satisfies
%%%%%%%%
\Align{
J^{(r)(k)}_T&=(k-1)\sum_{j=1}^{r}J^{(r)(k)}_{T-j}\ \brac{r+1\leq T}.
\label{eq:krec}
}
%%%%%%%%
The average meeting time is given as $\overline{T}_{\epsilon_r}=k^{r}$.
The largest root of the characteristic polynomial of the above recurrence relation, denoting it by $\lambda_{(r)(k)}$, becomes, for large $r$, as
%%%%%%%%
\Align{
\lambda_{(r)(k)}= k-(k-1)k^{-r}+{\cal O}(k^{-2r}).
}
%%%%%%%%
Following 
the same argument as in the binary case, 
the limiting normalized distribution asymptotically 
takes the form as 
%%%%%%%%
\Align{
\lim_{r\to\infty} m'_{\epsilon_r}(\tau)\sim \exp\brac{-\frac{k-1}{k}\tau}.
}
%%%%%%%%
We should emphasize 
that the exponent of the limiting exponential distributions depends on the number of symbols, 
while the \emph{recurrence time} has a simple exponential distribution, 
as far as the system is 
hyperbolic \cite{Hir:ETDS:93}.

%%%%%%%%%%%%%%%%%%%%%%%%%%%%%%%%
\section{General Bernoulli systems}
\label{sec:4}

Next, 
let us consider $(p,q)$-Bernoulli systems where $0<p<1,\ q=1-p$.
Let $X$ be the unit interval, $X=[0,1)$. 
The $(p,q)$-Bernoulli map is defined as 
%%%%%%%%
\Align{
F^{(p,q)}(x)=
\Cases{
x/p&\brac{0<x<p}\\
(x-p)/q&\brac{p<x<1}
}
.
\label{eq:pq_map}
}
%%%%%%%%
Denote a semi-infinite binary sequence as
$\bfm{s}=s_1s_2\cdots$, where $\ s_i\in\braC{0,1}$.
The map is equivalent to the shift with $(p,q)$-Bernoulli measure through the following equation,
%%%%%%%%
\Align{
x= \sum_{i=1}^{\infty}s_{i}f(\overline{s_{i}})\varphi_{\bf s}(i-1),
\label{eq:symbol_point}
}
%%%%%%%%
where $ f(s_i)=\Cases{p&(s_i=0)\\q&(s_i=1)}$ and
$ \varphi_{\bf s}(n)=\prod_{i=1}^{n}f(s_i)$ and $\overline{s_{i}}$ is the complement of $s_{i}$.
In spite of construction of the symbolic expression, the argument in the previous sections 
(mainly developed 
in \ref{app:A}) 
cannot be directly applied to this case 
since the probability to appear a given symbol sequence depends on its symbols and $p$ (and $q$) 
as well.
In order to describe the meeting time for $(p,q)$-Bernoulli systems, 
we start with 
the case in which $x_1$ is periodic and $x_2$ is a non-periodic generic orbit.

\subsection{The meeting time for periodic orbits}
\label{sec:4:1}

Suppose that $x_1$ is a periodic orbit of period $L$ and denote its symbol 
by $\bfm{t}$.
Obviously, $\bfm{t}$ is an $L$-periodic sequence, \ie $t_{i}=t_{i+L}$, 
and thus it is represented by the first $L$ symbols, $t_1\cdots t_L$.
Here, we put $\epsilon=\braccc{f(t_1)f(t_2)\cdots f(t_L)}^{r}$.
For $T\geq rL+1$, 
we assume that 
the meeting time distribution for the periodic orbit, denoting it by $m_{\epsilon,\bfm{t}}\brac{T}$, satisfies the following relation, 
%%%%%%%%
\Align{
\frac{m_{\epsilon,\bfm{t}}\brac{T+rL}}{\varphi_{\bfm{t}}(T+rL)}=C_{T+rL-1}\sum_{j=0}^{rL-1}\frac{m_{\epsilon,\bfm{t}}\brac{T+j}}{\varphi_{\bfm{t}}(T+j)},
\label{eq:pq_equ}
}
%%%%%%%%
where $C_{i}=f(\overline{t_i})/f(t_i)$.

The above relation is a generalization of \ref{app:A} to $(p,q)$-Bernoulli systems.
(Indeed it is equivalent to equation \eref{eq:rec} when $p=q=1/2$.) 
It should be noted however that 
the derivation in general cases is not a straightforward 
extension of $(1/2,1/2)$-Bernoulli systems, 
but is given via symbolic dynamics 
with $(p,q)$-Bernoulli measure \cite{AHYS:UP:10}. 
We here assume the relation \eref{eq:pq_equ} with some arguments 
which will be more closely discussed in \sref{sec:5}. 
This is justified 
since $x_1$ is periodic and we took appropriate $\epsilon$ correspondingly.

%%%% figure %%%%
\begin{figure}
\begin{center}
\includegraphics{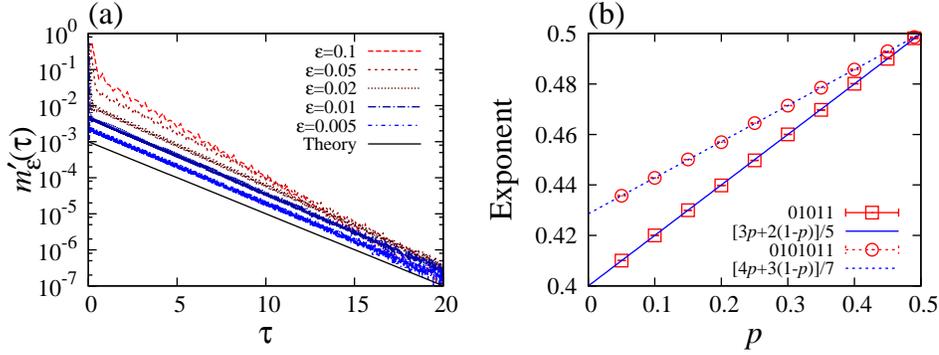}
\caption{
(a) The normalized meeting time distributions with respect to the periodic orbit $01011$ for the $(p,q)$-Bernoulli system with $p=0.3$.
The black line represents the expected one by equations \eref{eq:pq_dist_per} and \eref{eq:pq_dist_exp}.
(b) The exponent evaluated by fittings as a function of $p$.
The lines represent the exponent predicted by \eqref{eq:pq_dist_exp}.
Error bars are too small to be shown on the plot.
}
\label{figure2a_2b}
\end{center}
\end{figure}
%%%% figure %%%%

Let us consider a polynomial 
\Align{
\lambda^r-c\sum_{j=0}^{r-1}\lambda^j=0,
\label{eq:chara_poly_c}
} 
where $c$ is a constant.
In \ref{app:B}, we show that the largest root of the characteristic polynomial
behaves approximately as 
$\lambda_c=(1+c)-c(1+c)^{-r}+{\cal O}\brac{(1+c)^{-2r}}$ 
for large $r$.
If the right-hand side prefactor of the recurrence relation \eqref{eq:pq_equ} is constant $C_i=c$, 
the number given by the recurrence relation at most increases as $\lambda_c^T$.
Since we have $L$-periodic $C_i$, 
the asymptotic solution of equation \eref{eq:pq_equ} for large $T$ is given as 
$m_{\epsilon}\brac{T}/\varphi_{\bf t}(T)\sim \brac{\lambda_{C_1}\cdots\lambda_{C_L}}^{T/L}$
and then we obtain
%%%%%%%%
\Align{
m_{\epsilon,\bfm{t}}(T)\sim \brac{\Lambda_{1}\Lambda_{2}\cdots\Lambda_{L}}^{\frac{T}{L}},
\label{eq:dist_per}
}
%%%%%%%%
where $\Lambda_{i}=1-f(\overline{t_i})f(t_i)^r$.
Increasing $r$ (namely decreasing $\epsilon$), the factor $\Lambda_i$ should be normalized by $f(t_i)^r$, 
otherwise the distribution 
diverges.
We here assume that by replacing $\Lambda_i$ by $\Lambda_{i}^{f(t_i)^{-r}}$ and $T$ by $\tau$, 
the right-hand side of equation \eref{eq:dist_per} 
gives the asymptotic form of the normalized distribution $m'_{\epsilon,\bfm{t}}(\tau)$.
The normalized distribution is rewritten as
%%%%%%%%
\Align{
m'_{\epsilon,\bfm{t}}(\tau)&\sim \braccc{\Lambda_{1}^{f(t_1)^{-r}}\cdots\Lambda_{L}^{f(t_L)^{-r}}}^{\frac{\tau}{L}}.
\label{eq:pq_dist_nor}
}
%%%%%%%%
In the limit of large $r$ (small $\epsilon$), the distribution converges to an exponential 
function:
%%%%%%%%
\Align{
\lim_{r\to\infty}m'_{\epsilon,\bfm{t}}(\tau)\sim \exp\brac{-\alpha^{(p,q)}_{L}\tau}, 
\label{eq:pq_dist_per}
}
%%%%%%%%
where the exponent takes the average over the periodic orbit, namely
%%%%%%%%
\Align{
\alpha^{(p,q)}_{L}=\frac{1}{L}\sum_{i=1}^{L}f(\overline{t_i}).
\label{eq:pq_dist_exp}
}
%%%%%%%%
In \fref{figure2a_2b} we show the meeting time distributions obtained by numerical simulations of the $(p,q)$-Bernoulli map, 
showing good agreement with our theoretical prediction \eref{eq:pq_dist_per}.

In numerical simulations the average meeting time that normalizes the distribution 
is directly evaluated using equation \eref{eq:average} 
and is inversely proportional to $\epsilon$, namely $\overline{T}_\epsilon\approx1/2\epsilon$.
The results imply that the normalization adopted to derive equation \eref{eq:pq_dist_nor} is 
consistent with the numerical results.
In our derivations used to provide an asymptotic form of the distribution, 
one can calculate the average meeting time, 
but one sees that 
this average strongly depends on 
the parameters $(p,q)$ and the symbols of periodic orbits (as contrary to simple $\brac{1/k,\cdots,1/k}$-Bernoulli systems) \cite{AHYS:UP:10}.
The coincidence between the theoretical prediction and numerical observations implies 
that the normalization details 
are insensitive as far as one takes the limit $\epsilon \to 0$.

\subsection{The long period limit}
\label{sec:4:2}

\begin{figure}
\begin{center}
\includegraphics{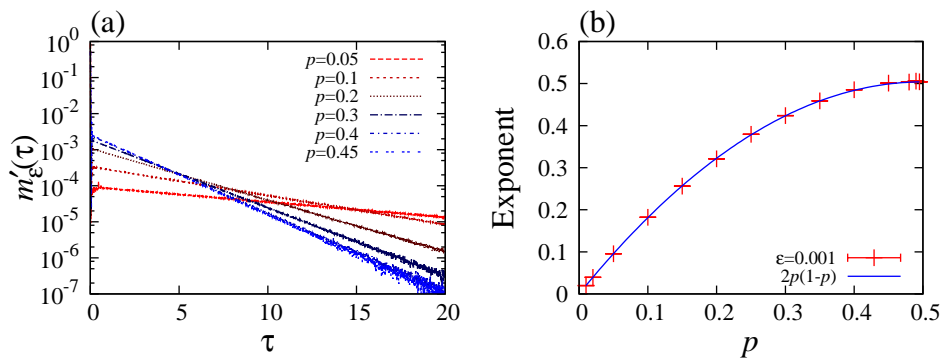}
\caption{
(a) The normalized meeting time distribution with $\epsilon=0.005$.
(b) The decaying exponent of meeting time distributions plotted as a function of $p$.
The curve represents the exponent predicted by \eqref{eq:pq_exponent}.
Error bars are too small to be visible on the plot.
}
\label{figure3a_3b}
\end{center}
\end{figure}

Here we consider the meeting time distribution for which $x_1$ is non-periodic.
We assume that the meeting time for non-periodic orbits is 
well approximated by the meeting time for long periodic orbits. 
Since the argument in the previous section concluded that 
the average with respect to the period gives the exponent, 
the distribution for non-periodic $x_1$ is exponential for which the exponent is given by the ergodic average.
The exponent for the long period limit 
is shown to be
%%%%%%%%
\Align{
\alpha^{(p,q)}&=\lim_{L\to\infty}\alpha^{(p,q)}_{L}=2pq.
\label{eq:pq_exponent}
}
%%%%%%%%
\Fref{figure3a_3b} shows the 
distributions obtained numerically for different $p$. 
Each exponent excellently agrees with the prediction of equation \eref{eq:pq_exponent} when $\epsilon$ is sufficiently small.

%%%% figure %%%%
%%%% figure %%%%

\subsection{$(p_1,\cdots,p_k)$-Bernoulli systems}
\label{sec:4:3}

\begin{figure}
\begin{center}
\includegraphics{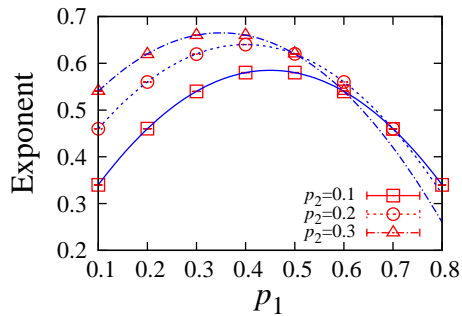}
\caption{
The exponent of the normalized meeting time for $(p_1,p_2,p_3)$-Bernoulli maps.
The curves represent the exponent predicted by \eqref{eq:k_exponent}.
Error bars are too small to be shown on the plot.
}
\label{figure4}
\end{center}
\end{figure}

The extension of the previous discussions to $(p_1,\cdots,p_k)$-Bernoulli systems is straightforward.
For $s_i\in\bbm{S}_k:=\braC{1,\cdots,k}$, define a function 
%%%%%%%%
\Align{
f^{(k)}(s_i)=\Cases{p_1&(s_i=1)\\&\vdots\\p_k&(s_i=k)},
}
%%%%%%%%
where $0<p_l<1$ and $\sum_{l=1}^{k}p_l=1$.
Then, the $(p_1,\cdots,p_k)$-Bernoulli map on the unit interval is defined as
%%%%%%%%
\Align{
F^{(k)}(x)=\Cases{
\brac{x-S_{0}}/p_1&
\brac{S_{0}<x<S_{1}}\\
&\vdots\\
\brac{x-S_{k-1}}/p_k&
\brac{S_{k-1}<x<S_{k}}
}
,
}
%%%%%%%%
where $ S_{s}=\sum_{l=1}^{s}p_l$.
Following the argument in \sref{sec:3:1},
the equation for a periodic orbit to satisfy the meeting time distribution 
is 
%%%%%%%%
\Align{
\frac{m_{\epsilon,\bfm{t}}\brac{T+rL}}{\varphi^{(k)}_{\bf t}(T+rL)}&
=C^{(k)}_{T+N-1}\sum_{j=0}^{rL-1}\frac{m_{\epsilon,\bfm{t}}\brac{T+j}}{\varphi^{(k)}_{\bf t}(T+j)},
}
%%%%%%%%
where ${\bf t}$ is 
the symbol of the periodic orbit
and $L$ is its period and
$\varphi^{(k)}_{\bf t}(n)=\prod_{i=1}^{n}f^{(k)}(t_i)$. 
The constant $C^{(k)}_i$ is given as
%%%%%%%%
\Align{
C^{(k)}_i&=\frac{1}{f^{(k)}(t_i)}
\sum_{s\in\bbm{S}_k,s\neq t_i}
f^{(k)}(s)=\frac{1-f^{(k)}(t_i)}{f^{(k)}(t_i)}.
}
%%%%%%%%
For the $(p_1,\cdots,p_k)$-Bernoulli map, 
the long period limit leads to the exponent of the normalized distribution as
%%%%%%%%
\Align{
\alpha^{(k)}=
\sum_{l=1}^{k}p_l(1-p_l).
\label{eq:k_exponent}
}
%%%%%%%%
Note that the obtained exponent is a natural extension of that for $(p,q)$-Bernoulli systems, and even for $\brac{1/k,\cdots,1/k}$-Bernoulli systems.
Numerical simulations show excellent agreement with our prediction \eqref{eq:k_exponent}, 
which is shown in \fref{figure4}.

%%%% figure %%%%
%%%% figure %%%%

%%%%%%%%%%%%%%%%%%%%%%%%%%%%%%%%
\section{Conclusions and discussion}
\label{sec:5}

In this paper, 
we have introduced the meeting time, the time interval for which two orbits approach each other within a given distance $\epsilon$.
We have shown that 
the distribution of the meeting time is exponential 
for $\brac{1/k,\cdots,1/k}$-Bernoulli systems.
In the limit of small $\epsilon$, the meeting time normalized by its average obeys an exponential distribution whose exponent is $(k-1)/k$.
For $(p_1,\cdots,p_k)$-Bernoulli systems, our analysis, based on periodic orbit approximations, 
predicts the exponent as $\alpha^{(k)}=\sum_{l=1}^{k}p_l(1-p_l)$.
This exponent varies in the range of $0< \alpha^{(k)}\leq (k-1)/k$ and 
is maximized when $p_1=\cdots=p_k=1/k$.

The meeting time distribution is similar to the recurrence time distribution, the latter  
being studied for several chaotic systems not necessarily hyperbolic ones \cite{CS:PRL:99,ZTRK:Cha:07}. 
We here point out two aspects on the relation between them.  

First, suppose that $x_1$ is a fixed point, then the meeting time is equivalent 
to time to return to the region of the $\epsilon$-neighborhood of $x_1$, 
and there is a rigorous proof that the recurrence time distribution is a simple exponential \cite{Hir:ETDS:93}.
The meeting time contains information how $x_1$ travels around in phase space.
In this sense, the meeting time captures more correlations among trajectories than the recurrence time.

%%%% figure %%%%
\begin{figure}
\begin{center}
\includegraphics{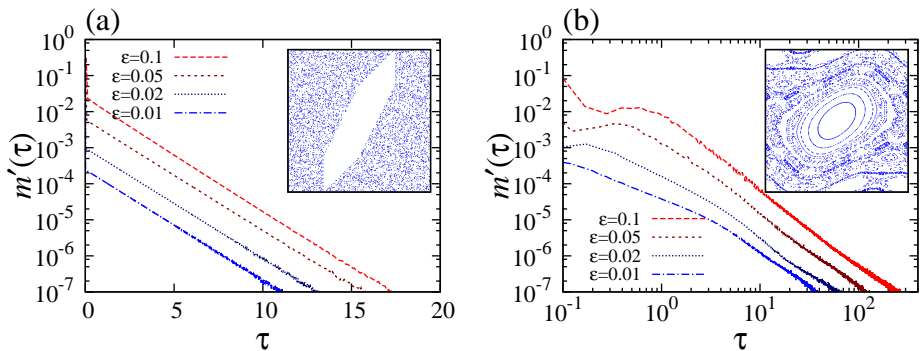}
\caption{
(a) The meeting time distribution for 
the piecewise linear map studied in \cite{AS:PRE:09} with $K=3$.
(semi-logarithmic plot).
The distribution for every $\epsilon$ fits to a simple exponential 
except small deviations due to statistical errors.
(b) The meeting time distribution for 
the standard map with $K=1$.
(double logarithmic plot).
We put two initial points on the largest chaotic component in the phase space. 
No quantitative changes of the distribution were found 
even though choosing any other initial points in the same component.
The distribution obeys a power law partially
in particular time regimes, while its time scales (and the power law exponents as well) 
vary as $\epsilon$ decreases.
}
\label{figure5a_5b}
\end{center}
\end{figure}
%%%% figure %%%%

Second, the meeting time distribution for the area-preserving map with mixed phase 
space behaves rather unexpectedly: 
as shown in \fref{figure5a_5b}(a),  
the meeting time distribution in case of the system with sharply-divided phase space is exponential 
for finite-value $\epsilon$. 
On the other hand, as shown in \cite{AMK:PRE:06,AS:PRE:09}, 
the recurrence time distribution for the system exhibits power law decay and the corresponding exponent 
can be derived theoretically. 
Furthermore, the meeting time distribution for a generic system whose phase space 
forms hierarchical mixtures of stable and chaotic regions cannot fit overall either to 
a simple function such as exponential or a power law (see \fref{figure5a_5b}(b)). 
As mentioned in introduction, the power law exponent 
of the recurrence time distribution for generic mixed systems is still an unsettled issue, 
but there exists, at least, a consensus that the distribution can be fitted by
power law decaying functions in a wide range \cite{WHK:PRL:02,CK:PRL:08}. 
In this way, the meeting time distribution for mixed systems 
makes a sharp contrast, which implies that these two measures capture 
different aspects of complex behavior in hierarchical phase space.

The origin of the intricate distribution of \fref{figure5a_5b}(b) is 
undoubtedly 
associated with complex hierarchical structures.
In order to study complex phase space, one 
may define the meeting time distribution in different ways: 
for a given point $z\in X$ and $\epsilon>0$, let $x$ be a point for which $d(z,x)<\epsilon$.
The meeting time for $x$ is defined as
%%%%%%%%
\Align{
T_{z,\epsilon}(x):=\inf\braC{n\in\bbm{N}~|~d\brac{F^{n}(z),F^{n}(x)}<\epsilon}.
}
%%%%%%%%
The distribution for $z$ is defined as 
%%%%%%%%
\Align{
M_{z,\epsilon}(T):=\mu\brac{\braC{x\in X~|~d(z,x)<\epsilon, T_{z,\epsilon}(x)=T}}, 
\label{eq:def_dist_z}
}
%%%%%%%%
where $\mu\brac{\cdot}$ is an appropriate measure.
The defined distribution $M_{z,\epsilon}(T)$ is a point-wise measure when one takes the limit $\epsilon\to0$, 
and thus it essentially depends on $z$.
The ergodic average of $M_{z,\epsilon}(T)$ with respect to $z$ is supposed to be equal to $m_{\epsilon}(T)$.
Investigating $M_{z,\epsilon}(T)$ by changing $z$ might be a key to study the case in which phase space is inhomogeneous.

As mentioned in section \ref{sec:4:1}, the derivation of \eqref{eq:pq_equ} requires 
the meeting time for symbolic dynamics.
The meeting time distribution for symbol sequences is defined as \eqref{eq:def_dist_z}, with the number $n$ of how many symbols coincide, 
instead of $\epsilon$ \cite{AHYS:UP:10}, denoting it by $M_{\bfm{s},n}(T)$ where $\bfm{s}$ is a given symbol sequence.
In general $M_{\bfm{s},n}(T)$ differs from $M_{z,\epsilon}(T)$ since 
the distance for symbolic dynamics is not necessarily equal to that for corresponding map systems.
If $z$ (and the corresponding symbol sequence $\bfm{t}$) is periodic, 
the two distribution are identical.
In fact, by taking $\epsilon=[f(t_1)f(t_2)\cdots f(t_L)]^{r}$, for a point $y$ and the corresponding symbol sequence ${\bf u}$, 
it holds that 
the first $rL$ symbols of ${\bf u}$, at least, coincide with those of ${\bf t}$ 
iff $|y-z|<\epsilon$ is satisfied.
The same statement holds for $F^i(z)$ (and its symbol expression) with 
the same $\epsilon$
and thus $M_{z,\epsilon}(T)=M_{\bfm{t},rL}(T)$.
What remains to be shown is the following relation 
%%%%%%%%
\Align{
m_{\epsilon,{\bf t}}(T)=\frac{1}{L}\sum_{i=0}^{L-1}M_{F^{i}(z),\epsilon}(T). 
}
%%%%%%%%
The distribution in the left-hand side is defined by \eqref{eq:def_dist} 
as the long time average of how often $x_2$ ``meets'' one of the periodic points.
The right-hand side is an average over the distribution with respect to each periodic point.
This relation is supposed to be justified 
if $x_2$ uniformly visits the $\epsilon$-neighborhood of one of the periodic points with the probability $1/L$.
These arguments validate that the meeting time distribution for periodic points satisfies \eqref{eq:pq_equ}.

Recall that the meeting time is the recurrence time for the cross product system, as mentioned in the last paragraph of \sref{sec:2}.
While the meeting time is a special case of the recurrence time, 
our numerical observations reveal that the recurrence for the cross product system, 
a couple of two identical dynamical systems, significantly differs from the one for the original system.
The recurrence time of a typical dynamical system, regardless of its dimensionality, 
has a simple exponential distribution in chaotic systems \cite{Hir:ETDS:93}, 
or a power law one in mixed systems \cite{CS:PRL:99}.
This in turn suggests that there exist dynamical systems whose
recurrence time distributions, the recurrence region being taken in a special way,
do not obey those in generic systems.

The meeting time that we defined for maps can be similarly considered in continuous-time dynamical systems as well.
Needless to say, the normalization and/or the small $\epsilon$ limit should be properly taken into account in slightly different ways.
Although we here do not discuss in detail, 
we expect that the meeting time distribution is exponential in fully-chaotic phase space.
If this is indeed the case, the exponent 
may have physical significance, 
or even the relationship to other physical quantities that characterize chaotic properties.

%%%%%%%%%%%%%%%%%%%%%%%%%%%%%%%%
\appendix
\section{Meeting time distributions for $\brac{\frac{1}{2},\frac{1}{2}}$-Bernoulli systems}
\label{app:A}

Let $X$ be $\braC{0,1}^{\bbm{N}}$ and denote a semi-infinite symbol sequence, ${\bf s}\in X$, as
%%%%%%%%
\Align{
{\bf s}=s_1s_2s_3\cdots,
}
%%%%%%%%
where $s_i\in\braC{0,1}$, 
and $F$ be the shift map, \ie $F({\bf s})=s_2s_3s_4\cdots$.
For ${\bf s},{\bf s}'\in X$, the distance is defined as
%%%%%%%%
\Align{
d({\bf s},{\bf s}')=\sum_{i=1}^{\infty}2^{-i}|s_i-s'_i|.
}
%%%%%%%%
Putting $\epsilon=\epsilon_r:=2^{-r}$ where $r$ is a positive integer, one can see that ${\bf s}$ meets ${\bf s}'$, namely $d({\bf s},{\bf s}')<\epsilon_r$, 
iff the first $r$ symbols of ${\bf s}$ and ${\bf s}'$ coincide, \ie $s_i=s'_i, i=1,\cdots,r$.

Taking account of the fact that every symbol, $s_i$ or $s'_i$, is either `0' or `1' with the probability $1/2$, 
the meeting time is given as follows:
let ${\bf s}$ be fixed without loss of generality and the first $r$ symbols of ${\bf s}'$ are those of ${\bf s}$ 
and for $i\geq r+1$ $s'_i$ is either $s_i$ or $\overline{s_i}$ where $\overline{s_i}$ is the complement of $s_i$.
Let us consider a symbol sequence, $t_{r+1}\cdots t_{r+n}$, where $t_{i}$ is either $s_{i}$ or $\overline{s_{i}}$.
For a given $t_{r+1}\cdots t_{r+n}$, 
let $[t_{r+1}\cdots t_{r+n}]$ denote the subset of ${\bf s}'$ such that $s'_{r+i}=t_{r+i},\ i=1,\cdots,n$ 
and call it as a sequence of length $n$.
The probability to appear a symbol sequence in $[t_{r+1}\cdots t_{r+n}]$ is $2^{-n}$, denoting it by $\mu\brac{[t_{r+1}\cdots t_{r+n}]}=2^{-n}$.

The meeting time distribution $m_{\epsilon_r}(T)$ is given by the probability of 
symbol sequences of ${\bf s}'$ such that, after $T$-times shift, the first $r$ symbols coincide with those of ${\bf s}$, 
namely the $r$ symbols after the $T$th symbol coincide $s'_{T+i}=s_{T+i}, i=1,\cdots,r$.
For $T=1$ the distribution is given by $[s_{r+1}]$ \ie $m_{\epsilon_r}(1)=\mu([s_{r+1}])=1/2$.
For $T\geq 2$ a symbol is a subset of $[\overline{s_{r+1}}]$.
For $2\leq T\leq r$, $m_{\epsilon_r}(T)=0$ since $\overline{s_{r+1}}$ is in the $r$ symbols to coincide.
For $T\geq r+1$, the distribution $m_{\epsilon_r}(T)$ is given by 
finite sequences of length $T$ with following two properties:
\Description{
\item[(i)] Any of the last $r$ symbols are not the complement $\overline{s_{i}}$
(the last $r$ symbols are expressed as $[\cdots s_{T+1}\cdots s_{T+r}]$).
\item[(ii)] There are no $r$-consecutive symbols of the non-complement expression before the last $r$ symbol sequence.
(the $r$ non-complement symbols  appear only in the last of the whole sequence, 
otherwise its meeting time is smaller than $T$.)
}
Let here $J_{T}^{(r)}$ denote the number of such sequences of length $T$.
One finds that $J_{r+1}^{(r)}=1$ since $[\overline{s_{r+1}}s_{r+2}\cdots s_{2r+1}]$ is the unique symbol sequence for $T=r+1$, 
and $J_{r+2}^{(r)}=1$, since $[\overline{s_{r+1}s_{r+2}}s_{r+3}\cdots s_{2r+2}]$ for $T=r+2$.
For $T\geq r+3$, consider the following sequence,
%%%%%%%%
\Align{
\overline{s_{r+1}}\underbrace{t_{r+2}\cdots t_{T-1}}\overline{s_{T}}s_{T+1}\cdots s_{T+r}.
}
%%%%%%%%
For $T\leq 2r+1$, the bracket part $t_{r+2}\cdots t_{T-1}$ allows complete binary combinations since its length is less than $r$, and therefore
 $J_{T}^{(r)}=2^{T-r-2}$.
 
The number sequence $J^{(r)}_T$, which is the combinatorial number of $t_{r+2}\cdots t_{T-1}$ 
containing no $r$-consecutive sequences of $\overline{s_i}$, 
satisfies the recurrence relation given as
%%%%%%%%
\Align{
J^{(r)}_T&=\sum_{j=1}^{r}J^{(r)}_{T-j}\ \brac{r+1\leq T}.
\label{eq:app_rec}
}
%%%%%%%%
This is simply justified as follows:
let us consider a set of the binary sequences $\{0,1\}$ of length $i$ which do not contain $r$-consecutive `0's.  
Denote this set by $B_{i}^{(r)}$ and $n_{i}^{(r)}=\# B_{i}^{(r)}$.
($n_{i}^{(r)}$ is obviously equivalent to $J_{i}^{(r)}$, by replacing $\overline{s_i}$ with `0'  and $s_i$ `1'.)
In order to construct $B_{i}^{(r)}$, let us start with the simplest case $r=2$:
the set of length 1 is $B_{1}^{(2)}=\{0,1\}$.
$B_{i}^{(2)}$ is generated from $B_{i-1}^{(2)}$ by the following two rules:
{\bf (a)} append `1' to every element of $B_{i-1}^{(2)}$ 
and {\bf (b)} append `0' to an element of $B_{i-1}^{(2)}$ if the last symbol of the element is not `0'.
Then, $B_{i}^{(2)}$ is recursively generated, e.g. $B_{2}^{(2)}=\{01,11,10\}$, $B_{3}^{(2)}=\{011,111,101,010,110\}$, $\cdots$.
The appending rules (a) and (b) lead the recurrence relation, $n_{i}^{(2)}=n_{i-1}^{(2)}+n_{i-2}^{(2)}$:
obviously $n_{i-1}^{(2)}$ represents the number of elements generated by (a).
The number of elements by (b) is $n_{i-2}^{(2)}$ 
since the elements of $B_{i-1}^{(2)}$ whose last symbol is not `0' are generated from elements of $B_{i-2}^{(2)}$ by (a).
For $r\geq3$, $B_{i}^{(r)}$ is generated from $B_{i-1}^{(r)}$ as well by (a) and the modified rule of (b):
{\bf (b')} append `0' to an element of $B_{i-1}^{(2)}$ if the last $r-1$ symbol sequence is not $(r-1)$-consecutive `0's.
(e.g. for $r=3$ $B_{2}^{(3)}=\{0,1\}$, $B_{2}^{(3)}=\{01,11,00,10\}$, $B_{3}^{(3)}=\{011,111,001,101,010,110,100\}$, $\cdots$.)
The number of elements for which the modified rule (b') is applied is $n_{i-2}^{(r)}+\cdots+n_{i-r}^{(r)}$ 
since an element of $B_{i-1}^{(r)}$ generated from $B_{i-2}^{(r)},\cdots,B_{i-r}^{(r)}$ by (a) 
contains at least one `1's in the last $r-1$ symbol sequence, that is indeed the requirement of the modified rule (b').
Thus, one obtains the recurrence relation
$n_{i}^{(r)}=\sum_{j=1}^{r}n_{i-j}^{(r)}$ and equation \eqref{eq:app_rec}.
\\

Before closing the Appendix, we make a remark on the case of $k$ symbols.
The settings for $k$-symbol cases are simply given, as described in \sref{sec:3:2}.
Since the complement of $s_i$ has $k-1$ candidates of the symbols, 
the recurrence relation which the numbers of the sequences that hold {\bf (i)} and {\bf (ii)} in $k$-symbol cases satisfy
has the factor $k-1$ in the right-hand side of equation \eref{eq:app_rec}.
(In other words, the rule (a) in $k$-symbol cases allows to append any of $\{1,2,\cdots,k-1\}$ 
so that the factor $k-1$ appears.)

%%%%%%%%%%%%%%%%%%%%%%%%%%%%%%%%
\section{The real largest root of equation \eqref{eq:chara_poly_c}}
\label{app:B}
Let us rewrite the left-hand side of equation \eqref{eq:chara_poly_c} as 
%%%%%%%%
\Align{
P_{r,c}(\lambda)=\lambda^r-c\frac{\lambda^r-1}{\lambda-1},
}
%%%%%%%%
where $c>0$ is a real constant and $r\geq1$ is an integer.
One can easily show 
$P_{r,c}(1+c)
=1$ 
and for $d>0$, 
%%%%%%%%
\Align{
P_{r,c}(1+c+d)&=(1+c+d)^r-c\frac{(1+c+d)^r-1}{c+d},\nonumber\\
&=[(1+c+d)^r-1]\frac{d}{c+d}+1>1. 
}
%%%%%%%%
One finds that the real largest root of the polynomial is smaller than $1+c$.

The derivative of $P_{r,c}(\lambda)$ at $\lambda=1+c$ is $P'_{r,c}(1+c)=\frac{(1+c)^r-1}{c}>0$ and is of order of $(1+c)^r$.
Therefore, for large $r$, the largest root is supposed to be close to $1+c$ and can be obtained by perturbative calculations.
Let us rewrite the polynomial, by the expansion around $1+c$,  as 
%%%%%%%%
\Align{
P_{r,c}(1+c+x)=P_{r,c}(1+c)+&P'_{r,c}(1+c)x+\frac{1}{2!}P''_{r,c}(1+c)x^2+\cdots\nonumber\\
&+\frac{1}{r!}P^{(r)}_{r,c}(1+c)x^r,
}
%%%%%%%%
and 
expand $x$ by the power series of $(1+c)^{-r}$ as
%%%%%%%%
\Align{
x=a_1(1+c)^{-r}+a_2(1+c)^{-2r}+\cdots.
}
%%%%%%%%
Then, the expansion of the polynomial becomes as
%%%%%%%%
\Align{
P_{r,c}(1+c+x)&=1+\frac{(1+c)^r-1}{c}[a_1(1+c)^{-r}+a_2(1+c)^{-2r}+\cdots],\nonumber\\
&+\frac{1}{2}P''_{r,c}(1+c)[a_1(1+c)^{-r}+a_2(1+c)^{-2r}+\cdots]^2+\cdots,\nonumber\\
&=1+\frac{a_1}{c}+{\cal O}((1+c)^{-r}).
}
%%%%%%%%
The root is obtained by calculating $a_i$ such that every order of $(1+c)^{-r}$ becomes zero.
One obtains $a_1=-c$ for the lowest order and the root as $(1+c)-c(1+c)^{-r}+\cdots$.

\section*{References}
%%%%
%\bibliographystyle{iopart-num}
%\bibliography{chaos_01,math_01}

\begin{thebibliography}{10}
\expandafter\ifx\csname url\endcsname\relax
  \def\url#1{{\tt #1}}\fi
\expandafter\ifx\csname urlprefix\endcsname\relax\def\urlprefix{URL }\fi
\providecommand{\eprint}[2][]{\url{#2}}

\bibitem{LL:book:83}
Lichtenberg A~L and Lieberman M~A 1983 {\em Regular and chaotic motion\/} (New
  York: Springer-Verlag)

\bibitem{Gas:book:98}
Gaspard P 1998 {\em Chaos, Scattering and Statistical Mechanics\/} (Cambridge:
  Cambridge University Press)

\bibitem{Sko:LNP:10}
Skokos C 2010 The lyapunov characteristic exponents and their computation {\em
  Dynamics of Small Solar System Bodies and Exoplanets\/} ({\em Lecture Notes
  in Physics\/} vol 790) ed Souchay J~J and Dvorak R (Springer Berlin /
  Heidelberg) pp 63--135

\bibitem{CFS:book:82}
Cornfeld I~P, Fomin S~V and Sinai Y~G 1982 {\em Ergodic theory\/} (Berlin:
  Springer)

\bibitem{Ott:book:02}
Ott E 2002 {\em Chaos in dynamical systems\/} (Cambridge: Cambridge University
  Press)

\bibitem{CS:PRL:99}
Chirikov B~V and Shepelyansky D~L 1999 {\em Phys. Rev. Lett.\/} {\bf 82}
  528--531

\bibitem{ZPRTK:PRE:07}
Zou Y, Paz\'o D, Romano M~C, Thiel M and Kurths J 2007 {\em Phys. Rev. E\/}
  {\bf 76} 016210

\bibitem{ZTRK:Cha:07}
Zou Y, Thiel M, Romano M~C and Kurths J 2007 {\em Chaos\/} {\bf 17} 043101

\bibitem{JKA:EPL:10}
Johansson M, Kopidakis G and Aubry S 2010 {\em Europhys. Lett.\/} {\bf 91}
  50001

\bibitem{Hir:ETDS:93}
Hirata M 1993 {\em Ergod. Theory Dyn. Syst.\/} {\bf 13} 533--556

\bibitem{HSV:CMP:99}
Hirata M, Saussol B and Vaienti S 1999 {\em Comm. Math. Phys.\/} {\bf 206}
  33--55

\bibitem{WHK:PRL:02}
Weiss M, Hufnagel L and Ketzmerick R 2002 {\em Phys. Rev. Lett.\/} {\bf 89}
  239401

\bibitem{CK:PRL:08}
Cristadoro G and Ketzmerick R 2008 {\em Phys. Rev. Lett.\/} {\bf 100} 184101

\bibitem{Ven:PRL:09}
Venegeroles R 2009 {\em Phys. Rev. Lett.\/} {\bf 102} 064101

\bibitem{Mil:AMM:60}
Miles E~P 1960 {\em Amer. Math. Monthly\/} {\bf 67} 745--752

\bibitem{AHYS:UP:10}
Akaishi A, Hirata M, Yamamoto K and Shudo A to be submitted

\bibitem{AMK:PRE:06}
Altmann E~G, Motter A~E and Kantz H 2006 {\em Phys. Rev. E\/} {\bf 73} 026207

\bibitem{AS:PRE:09}
Akaishi A and Shudo A 2009 {\em Phys. Rev. E\/} {\bf 80} 066211

\end{thebibliography}
%%%%

\providecommand{\newblock}{}

\end{document}